\documentclass[]{spie}  %>>> use for US letter paper
\newcommand{\urss}[1]{\ensuremath{_{\mathrm{#1}}}}

\usepackage{amsmath,amsfonts,amssymb}
\usepackage{subfigure}
\usepackage{graphicx}
\usepackage{upgreek}
\usepackage{xcolor}
\usepackage[colorlinks=true, allcolors=blue]{hyperref}
\usepackage{lineno}

\title{System performance of a cryogenic test-bed for the time-division multiplexing readout for NewAthena X-IFU}

\author[1]{D.~Vaccaro}
\author[2]{J.~van~der~Kuur}
\author[1]{P.~van~der~Hulst}
\author[1]{T.~Vos}

\author[1]{M.~de~Wit}
\author[1]{L.~Gottardi}
\author[1]{K.~Ravensberg}
\author[1]{E.~Taralli}

\author[3,4]{J.~Adams}
\author[4]{S.~R.~Bandler}
\author[5]{D.~Bennett}
\author[4]{J.~A.~Chervenak}
\author[5]{W.~B.~Doriese}
\author[5,6]{M.~Durkin}
\author[5,6]{J.D.~Gard}
\author[5]{C.~Reintsema}
\author[3,4]{K.~Sakai}
\author[4]{S.~J.~Smith}
\author[5,6]{J.N.~Ullom}
\author[3,4]{N.~A.~Wakeham}

\author[1,7]{J.W.A.~den~Herder}
\author[2]{B.~Jackson}
\author[1]{P.~Khosropanah}
\author[1,8]{J-R.~Gao}
\author[2]{P.~Roelfsema}
\author[1]{A.~Simionescu}

\affil[1]{NWO-I/SRON Netherlands Institute for Space Research, 2333CA Leiden, Netherlands}
\affil[2]{NWO-I/SRON Netherlands Institute for Space Research, 9747AD Groningen, Netherlands}
\affil[3]{CSST University of Maryland Baltimore County, Baltimore, MD21250 USA}
\affil[4]{NASA Goddard Space Flight Center, Greenbelt, MD 20771, USA}
\affil[5]{NIST, 325 Broadway, Boulder, CO, 80305-3328, USA}
\affil[6]{Department of Physics, University of Colorado, Boulder, CO, 80309, USA}
\affil[7]{Universiteit van Amsterdam, Science Park 904, 1090GE Amsterdam, Netherlands}
\affil[8]{Optics Group, Delft University of Technology, Delft, 2628CJ, Netherlands}

\authorinfo{Corresponding author: d.vaccaro@sron.nl}

\begin{document} 
\maketitle
\pagenumbering{arabic}

\begin{abstract}
\textcolor{black}{The X-ray Integral Field Unit (X-IFU) is an instrument of ESA's future NewAthena space observatory,  with the goal to provide high-energy resolution ($<$ 4 eV at X-ray energies up to 7 keV) and high-spatial resolution (9") spectroscopic imaging over the X-ray energy range from 200 eV to 12 keV, by means of an array of about 1500 transition-edge sensors (TES) read out via SQUID time-division multiplexing (TDM).  A TDM-based laboratory test-bed has been assembled at SRON, hosting an array of $75\times 75\ \upmu$m$^2$ TESs that are read out via 2-column $\times$ 32-row TDM.  A system component that is critical to high-performance operation is the wiring harness that connects the room-temperature electronics to the cryogenic readout componentry.  We report here on our characterization of such a test-bed, whose harness has a length close to what envisioned for X-IFU,  which allowed to achieve a co-added energy resolution at a level of 2.7~eV FWHM at 6~keV via 32-row readout. In addition, we provide an outlook on the integration of TDM readout into the X-IFU Focal-Plane Assembly Development Model.}
\end{abstract}


\begin{quotation}
This paper has been accepted for publication in \textit{Journal of Astronomical Telescopes, Instrumentation and Telescopes}.
\end{quotation}

\section{Introduction}\label{intro}

NewATHENA (Advanced Telescope for High-Energy Astrophysics) is ESA's next X-ray astronomical observatory that will study the Hot and Energetic Universe\cite{heu} from an L1 Sun-Earth orbit, with a launch foreseen for 2037. On board the satellite there will be two instruments: the Wide Field Imager (WFI) designed for imaging surveys in a 40'$\times$40' field of view and 0.2~keV to 15~keV band with DEPFET pixels, and the X-ray Integral Field Unit \cite{xifu} (X-IFU) optimised for imaging spectroscopy. 

The X-IFU will make use of $\approx$1500 transition edge-sensors (TES) micro-calorimeters \cite{gottardites},  organized in a hexagonal array covering a field of view of 4' with a pixel size of $317~\upmu$m corresponding to 5'' PSF angular resolution. The sensors will be cooled down to a base temperature of 50~mK by means of a multi-stage Adiabatic Demagnetization Refrigerator (ADR) and are designed to operate in the soft X-ray energy band, between 200~eV and 12~keV,  with a predicted energy resolution at instrument level better than 4~eV for energies up to 7~keV.

To comply with the stringent requirements of mass, volume and heat load for the cryogenic stages of the instrument,  a multiplexing readout scheme is necessary. Time-division multiplexing (TDM)\cite{tdm} has been selected, as it is a mature technology routinely used in ground-based X-ray and $\upgamma$-ray spectrometers, such as the SSRL (Stanford Synchrotron Radiation Lightsource) beamlines at SLAC (USA) and recently in the Micro-X sounding rocket \cite{microx}.

In TDM readout, the TESs are continuously dc-biased and are each coupled to a first stage SQUID (SQ1), activated one at a time via a flux-actuated switch (FAS). SQ1 signals are amplified by a SQUID series array (SSA) and digitized by room temperature electronics. Each SQ1 represents a ``row" and the FASs are activated by sequential square wave signals, or row-addressing signals (RAS), generated by the digital readout electronics (DRE).  The SQ1s are hosted on ``MUX" chips, together with the shunt resistance $R\urss{sh}$ to voltage bias the TES and Nyquist inductors $L\urss{Ny}$ to limit the pixel bandwidth. Each row is active for a certain ``row time",  all the $N$ rows are read out within one ``frame" and the $M$ columns are read out in parallel.  \textcolor{black}{An explicative depiction of the operation of a TDM readout for micro-calorimeters can be found in Figure~8 of the recently published review \emph{``Signal Readout for Transition-Edge Sensor X-ray Imaging Spectrometers"}}\cite{tdm}. For X-IFU, a 48-row $\times$ 32-column TDM readout is envisaged, with row time of 160~ns.

SRON is responsible for the design, development and delivery of X-IFU's Focal Plane Assembly (FPA). The first iteration of the FPA Development Model (FPA-DM) was employing a TES array\cite{tessron} and a frequency-domain multiplexing (FDM)\cite{fdmdemo} technology, both developed at SRON. It was successfully used to validate its thermo-mechanical and magnetic environment. A second iteration of the FPA-DM, called DM1.1\cite{fpadm}, is foreseen before upscaling to the FPA Engineering Model, which will employ a TDM architecture.  For the assembly and test campaign of the DM1.1, expected to start in Q2 2025, SRON will receive key components from several external partners: the TES array and the digital room temperature electronics (DRE) from NASA-GSFC (USA), the MUX chips from NIST (USA) and the warm front-end electronics (WFEE) from APC-CNRS (France).

In late 2022, our team assembled components from SRON, NIST and NASA-GSFC into a new cryogenic test-bed at SRON, as described in Section~\ref{setup}. The motivation was to transfer TDM technology from the development laboratory to a more flight-like configuration ahead of its implementation in the DM1.1. Driven by the NewAthena spacecraft geometry, the flight-like TDM system will require a significantly longer wiring harness,  linking the room-temperature electronics to the cryogenic componentry, than what is typically used in laboratory TDM systems. 

\textcolor{black}{Such a harness is critical for several aspects of the system operation, including crosstalk,  low-frequency signal transfer from the SSA SQUID to the low-noise amplifier (LNA) at the room temperature electronics, the noise contribution of the harness resistance and the open-loop bandwidth of the signal transfer.  Parameters such as harness length and termination on both the cold and the warm part of the readout have an impact on the aforementioned aspects in the system. }

This new test-bed at SRON uses a longer wiring harness than laboratory TDM systems.  An upgrade in late 2023 introduced a re-designed version of the long harness (see Section~\ref{cryosetup}).  Our main recent goal was to reproduce in this new system the nominal spectral performance of the previous laboratory systems: 2.8~eV FWHM or better energy resolution at 6~keV under 2-column~$\times$~32-row multiplexing.

In this paper we report on the experimental activities performed with such a cryogenic test-bed at SRON.  In Section~\ref{setup} we describe both the hardware and software composing the experimental setup. In Section~\ref{char} we describe the characterization tests performed to optimize and debug the system. In Section~\ref{demo} we show the performance demonstration, in agreement with the nominal specifications. Finally, in Section~\ref{concl} we summarize our results and give a prospect for future activities.

\section{Description of the setup}\label{setup}

For ease of reading, we subdivide the components of the test-bed into three categories: cryogenic assembly, room temperature electronics and control software. 

\subsection{Cryogenic assembly}\label{cryosetup}

The cryogenic assembly is mounted in a Leiden Cryogenics dilution refrigerator (model CF-CS81-400), having a nominal cooling power of 400~$\upmu$W at 120 mK.  Typical base temperature of the mixing chamber during full-load operation, with multiple X-ray TES micro-calorimeter experiments assembled in the refrigerator, is 18~mK. The TDM system is depicted in Figure~\ref{system} and it consists of the following components.

\subsubsection*{50 mK snout}

\begin{figure}[h]
\centering
\includegraphics[width=1\textwidth]{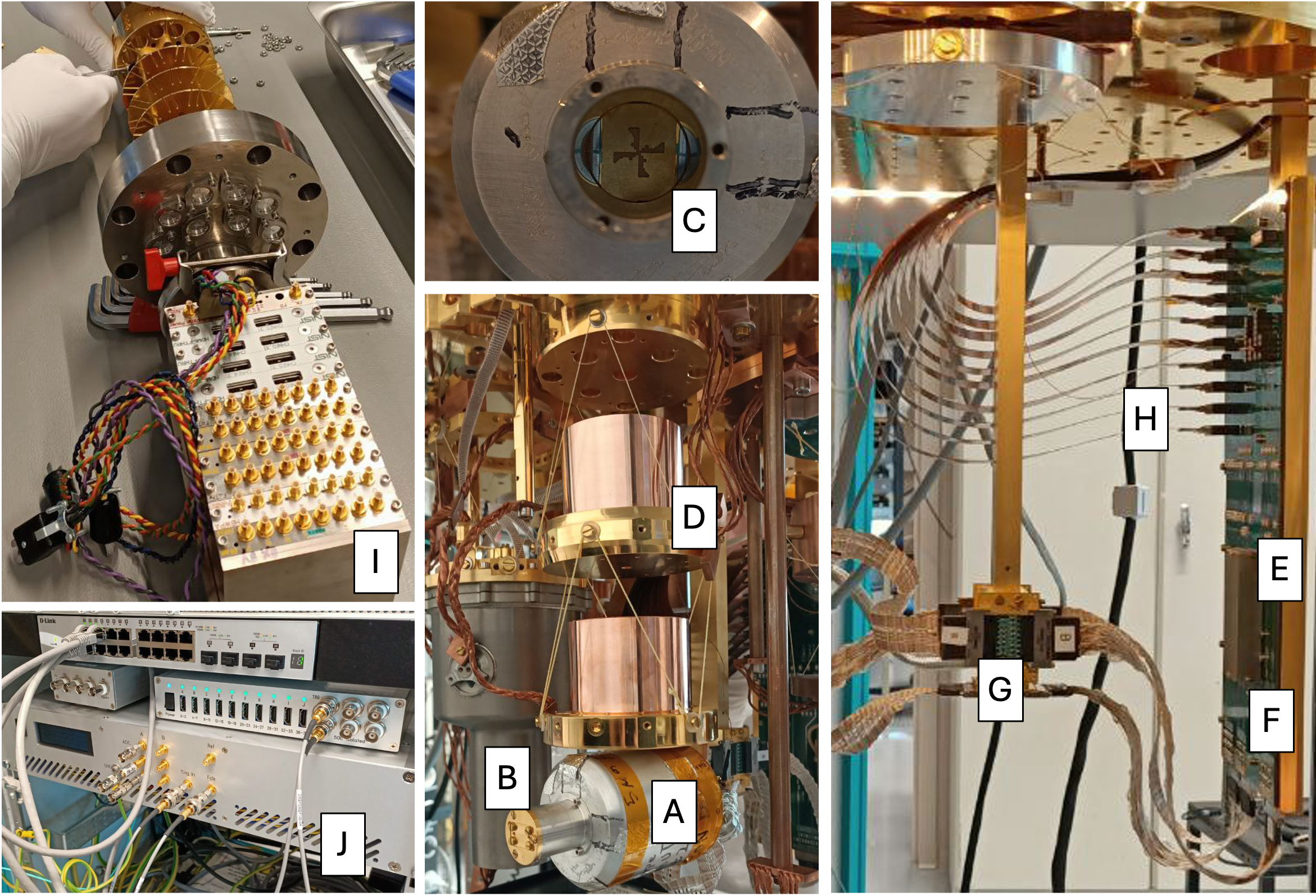}
\caption{Cryogenic components of the setup installed in the dilution refrigerator: 50 mK snout in Al superconducting shield (A),  Fe-55 X-ray source mounted on a Al chimney in front of the TES array (B), Cu aperture mask placed in between the TES array and the X-ray source (C), two-stage Kevlar suspension system (D), 4 K board (E), SSA SQUIDs (F), inter-thermal stage boards (G) for the NbTi braided looms connecting to the 4 K to the 50~mK snout, flex harness (H), Tower (I) during system assembly,  DRE (J) mounted in an electronics rack.}\label{system}
\end{figure}

A ``50 mK snout" \cite{snout} hosts the TES \cite{nasates} and MUX chips \cite{xtalkdurkin} ([A] in Figure~\ref{system}). The detector array, provided by NASA-GSFC, is a kilo-pixel chip with each TES made of $75 \times 75~\upmu$m$^2$ Mo/Au bilayer and Bi/Au absorber, with $R\urss{N} = 9.55$~m$\Omega$, \textcolor{black}{$C = 1.1$~pJ/K at 100 mK, $G = 84.7$~pW/K at $T\urss{C}$},  $T\urss{C} = 92.5$~mK. The MUX chips, manufactured at NIST, are of design ``mux20a"\cite{mux20a}, in which each SQ1 is a two-SQUID series array. Accompanying ``interface" chips, which contain the circuit elements for the TES-bias loop, are designed with $L\urss{Ny} = 640$~nH and $R\urss{sh} = 69~\upmu\Omega$. 

Both the TES array and MUX chips are different from the baseline design for X-IFU and are not optimized for the X-IFU level of performance, with an average single pixel energy resolution of $\approx 2.4$~eV at 6~keV: nevertheless,  these components are sufficient to allow for the technological transfer activity mentioned in Section~\ref{intro}.

The chips are enclosed in an aluminium casing to shield the TES and MUX chips from the influence of external magnetic fields. A Fe-55 radioactive source is mounted on a chimney on the aluminium casing in front of the TES array ([B] in Figure~\ref{system}), to illuminate the detectors with photons of 6~keV energy, exploiting the MnK$\upalpha$ spectral complex, used to characterize their energy resolution.  A copper aperture mask ([C] in Figure~\ref{system}) is placed between the TES array and the X-ray source, so that only the detectors connected to the readout are able to see the photons, to minimize thermal disturbances.

A magnetic field offset can be locally applied to the TES array by means of a superconducting coil placed directly below the detector chip.

The snout is suspended from the mixing chamber of the refrigerator using a two-stage Kevlar vibration isolation system \cite{gottardikevlar} ([D] in Figure~\ref{system}) to prevent power dissipation in the TES array due to mechanical contributions in the system and mitigate low-frequency contributions to the detectors' noise from such sources,  like the Pulse Tube. Thermal anchoring to the mixing chamber is performed via oxygen-free high-conductivity copper braids. Temperature is monitored via a ruthenium oxide thermometer and controlled via a PID (Proportional-Integral-Derivative) loop through a 750~$\Omega$ heater installed on the snout's mechanical support. An operating temperature of 50~mK is set and the typical RMS temperature stability achieved is at the level of 1~$\upmu$K.

\subsubsection*{4 K board and inter-stage harness}

The ``4 K board" ([E] in Figure~\ref{system}) hosts the SA20a series SQUID array (SSA), $i.e.$ the second-stage amplifiers for the TES signals.  Both the SSA and board assembly have been manufactured at NIST. The SSA chip is mounted within casing for magnetic shielding ([F] in Figure~\ref{system}), made of an external layer of niobium and an internal layer of cryogenic mu-metal \cite{reintsemanb}. The purpose of this board is as well to provide signal routing and filtering. A number of utility connections are also available, $e.g.$ for control of the superconducting magnet coil.

The board is mounted on a gold-plated copper mechanical support located below the mixing chamber, thermalised to the refrigerator's 3 K plate via a copper rod passing through sequential clearshots in the refrigerator's still and heat exchanger plates.  A Kevlar support structure is devised on top of the mixing chamber to dampen the mechanical vibrations of the board, given the weight of its mechanical support structure and the length of the rod.

This structure is motivated by necessity to limit the physical distance between the SSA and the MUX chips, to allow for sufficient interstage bandwidth between the two SQUID amplification stages, with the cut-off frequency $f\urss{3dB,interstage}$ given by $R\urss{dyn,SQ1}/(L\urss{in,SSA}+L\urss{stray})$, with $L\urss{in,SSA}$ the input inductance of the SSA, $R\urss{dyn,SQ1}$ the dynamic output resistance of the SQ1 (or MUX SQUID) and $L\urss{stray}$ the inductive contributions coming from the 50~mK to 3~K harness, PCB striplines, bonding wires, etc. Given the general requirement $f\urss{3dB,interstage} \geq 12$~MHz (for a row time of 160~ns) and typical values of $L\urss{in,SSA} \approx 100$~nH and $R\urss{dyn,SQ1} \approx$~15~$\Omega$, practical implementations require such harness length to be at a level of 30~cm. In this test-bed, such a harness is constituted by niobium-titanium twisted pair looms, with a thermal break in between via inter-stage filter cards ([G] in Figure~\ref{system}) thermally anchored to the heat exchanger plate ($\approx$~50~mK).

\subsubsection*{Flex harness}\label{secflex}

The connection between the ``4 K board" and the warm analog electronics (``Tower", see Subsection~\ref{warm}) is performed via the flex harness ([H] in Figure~\ref{system}). There are a total of 10 flexes, each constituted by eight supply and return conductor pairs, as diagrammed in Figure~2. The flexes are stacked and thermalised via dedicated heatsink plates at 3~K and 50~K. The interface of the flexes with the Tower electronics is described in Subsection~\ref{warm}.

\begin{figure}[h]%
\centering
\includegraphics[width=0.9\textwidth]{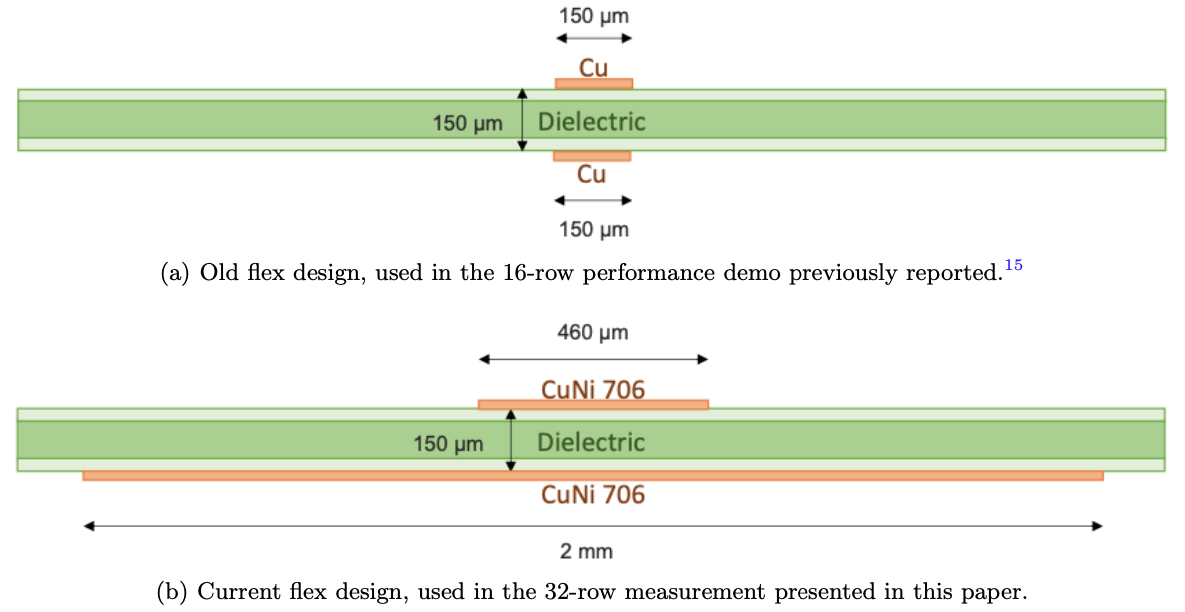}
\caption{Cross-section of the flex harness, comparing between the old (a) and current (b) flex design.  The direction of current flow along the conductors is into and out of the plane of the paper.  The pair pitch between adjacent wire pairs (not shown here) is 2.54~mm. The differences stand in the conductor material, symmetry between supply and return line and their individual width.}\label{flex}
\end{figure}

The flex harness used for the measurements reported in this paper differs from the one from our previous report \cite{srontdm}. Hereinafter, we refer to the previous design as ``old flex" and the current design as ``current flex". Figure~\ref{flex} shows cross-sectional diagrams to illustrate the differences between the two designs. 

For both designs, the flex length is at a level of  1.5~m, significantly longer than the standard $\approx$~40~cm used in ground-based TDM systems. The motivation for this difference was the reproduction of the nominal performance of a TDM system in an SRON dilution refrigerator, where the distance between the 3~K stage and room-temperature is larger than in the adiabatic demagnetization refrigerators (ADR) available at NIST and NASA-GSFC. Additionally, this is seen as an intermediate characterization of the system with a harness length closer to the 2.3~m required for X-IFU.

\textcolor{black}{The old flex design employs symmetrical supply and return conductor pairs, constituted by 150~$\upmu$m Cu lines with 2.54~mm pair pitch. }The dimensioning of the signal and return lines for the old flex design (Figure~\ref{flex}a) was chosen to keep the electrical resistance below 20~$\Omega$ to be compliant with the requirements for the electronics, as well as limiting the capacitance between signal and return lines to prevent bandwidth loss. This however resulted in a larger cable inductance which enhanced electrical crosstalk in the system, inducing large electrical oscillations (``ringing") with large amplitude, ultimately affecting the system performance already in first validation tests at NIST. 

This led to the new design of the signal and return lines (Figure~\ref{flex}b), similar to the standard dimensioning for the 40~cm length.  \textcolor{black}{In this case, two CuNi 706 conductor tracks are asymmetrical, with a 460 $\upmu$m wide supply line and a 2~mm wide return line, still with a 2.54~mm pair pitch. }Warm bench test measurements done at NIST showed that the level of electrical crosstalk in the harness was reduced by a factor of 20, which significantly reduced the amplitude of the ringing observed in the old flex. The wider signal track resulted however in a larger capacitance, which caused a reduction in harness bandwidth. The consequence of this is discussed in Section~\ref{char}.

\subsection{Room temperature electronics}\label{warm}

\subsection*{Analog front-end electronics}

The first stage of warm electronics is represented by the ``Tower" \cite{reintsemanb}, ([I] in Figure~\ref{system}) developed at NIST. The Tower is an ensemble of modular boards each assigned to a specific function: the amplification of the signal coming from the SSA via a low-noise amplifier\cite{nistlna} (LNA) and transmission to the ADC input of the DRE, the biasing via DAC circuitry to the quiescence operating point of TES, SQ1 and SSA, connection to the feedback circuits for the SSA (flux offset) and SQ1 (FLL) and routing of the row-addressing signals (RAS) coming from the DRE.

The control of the bias boards is performed via a serial protocol, executed from the measurement computer via an optical fiber connection, whereas the connections from the Tower to the DRE are via high-bandwidth HDMI cables for the RAS and BNC cables for the ADC and feedback signals.

\subsection*{Digital readout electronics}

The DRE \cite{kazudre} is developed at NASA-GSFC and is composed of two elements: the Column Box and the Row Box ([J] in Figure~\ref{system}). The Row Box generates the RAS to subsequently activate the rows of each column. Synchronicity between each row's activation window (row time) and ADC sampling performed by the Column box is ensured via several clocks (frame clock, line clock) generated from a master clock at 245.76 MHz. The Column box furthermore handles the digital feedback for the FLL as well as the DAQ back-end for data streaming.

\subsection{Control software}

Both the DRE from NASA-GSFC and the Tower from NIST have their own custom software package for measurement and control.  The Tower software is used to provide bias to the SSA and SQ1 SQUIDs and to the TES array, while the DRE software handles all the TDM control aspects, including row-timing and FLL settings, as well as measurement settings.

Both softwares are based on a graphical user interface that allow control and monitor of the readout parameters, as well as data logging for $e.g.$ X-ray data acquisition. To allow for automation, in view of the X-IFU FPA-DM1.1 test campaign, a dedicated control software has been developed at SRON.

This ``TES-TDM Readout Software" is built on SRON's generic ground support software platform. It is designed to be scalable to a high number of readout channels, with components that require high performance implemented in C++. In addition to automated measurement scripts, a text-based user interface is provided.

\section{System characterization}\label{char}

\subsection{EMI optimization}

To prevent ground loops,  a so-called ``star" grounding scheme is used, where the top of the cryostat is used as ground reference. All measurement electronics are grounded via thick copper braids to such reference.  Electronics modules, including the Column and Row Boxes, are powered on via an isolation transformer to decouple the ground reference of the TDM to the ground of the wall socket. Dedicated wall sockets in the laboratory,  connected to separated electrical units in the building, are used for powering on the readout electronics of cryogenic setups. The top of the cryostat is then connected to a ``clean" ground reference in the laboratory via a thick copper grounding cable. Furthermore, to electrically decouple the connections from the Tower and the Column box to the measurement computer, optical fibers are used.

Despite these precautions, strong noise lines at 50 Hz and higher harmonics have been initially observed in SQUID noise measurements. This was identified as due to electromagnetic interference (EMI) in the loop constituted by the $\sim$~3~m HDMI and BNC cables connecting the DRE to the Tower. In fact, it was observed that bundling and intertwining the cables together, to minimize the area of the loop and reject common-mode noise, significantly reduced the amplitude of such noise lines.  Finally, enclosing the cable bundle in aluminium foil almost completely removed the 50 Hz noise lines.  High-frequency (several kHz) lines, thought to be spurs from the DRE, remained present in the system. However, their frequency is sufficiently higher with respect to the $\approx$~250~Hz of the detector's electrical bandwidth at the quiescence operating point to consider their impact negligible. 

\subsection{Magnetic field offset}

As is customary for TES X-ray micro-calorimeter setups, to achieve the optimal spectral performance from the system, the behavior of the detectors under an external magnetic field has to be characterized. This is because the TES critical current is magnetic field dependent, as the system formed by the TES bilayer and the niobium leads constitutes an SS'S Josephson junction. It is typically observed that the best operation is obtained when no magnetic fields perpendicular to the bilayer are present.

These fields might be caused by external sources such as pulse tubes, Earth magnetic field, EMI from laboratory equipment, etc.  Furthermore, TESs under dc-bias suffer from the effects of self-field, $i.e.$ the magnetic field generated by the loop made by the leads and the TES itself \cite{smithb}. Such magnetic field adds to the residual field at the TES array and is bias point dependent. The external magnetic fields are typically compensated with a superconducting coil mounted just below the TES array,  as also envisaged for X-IFU.

\begin{figure}[h]%
\centering
\includegraphics[width=0.9\textwidth]{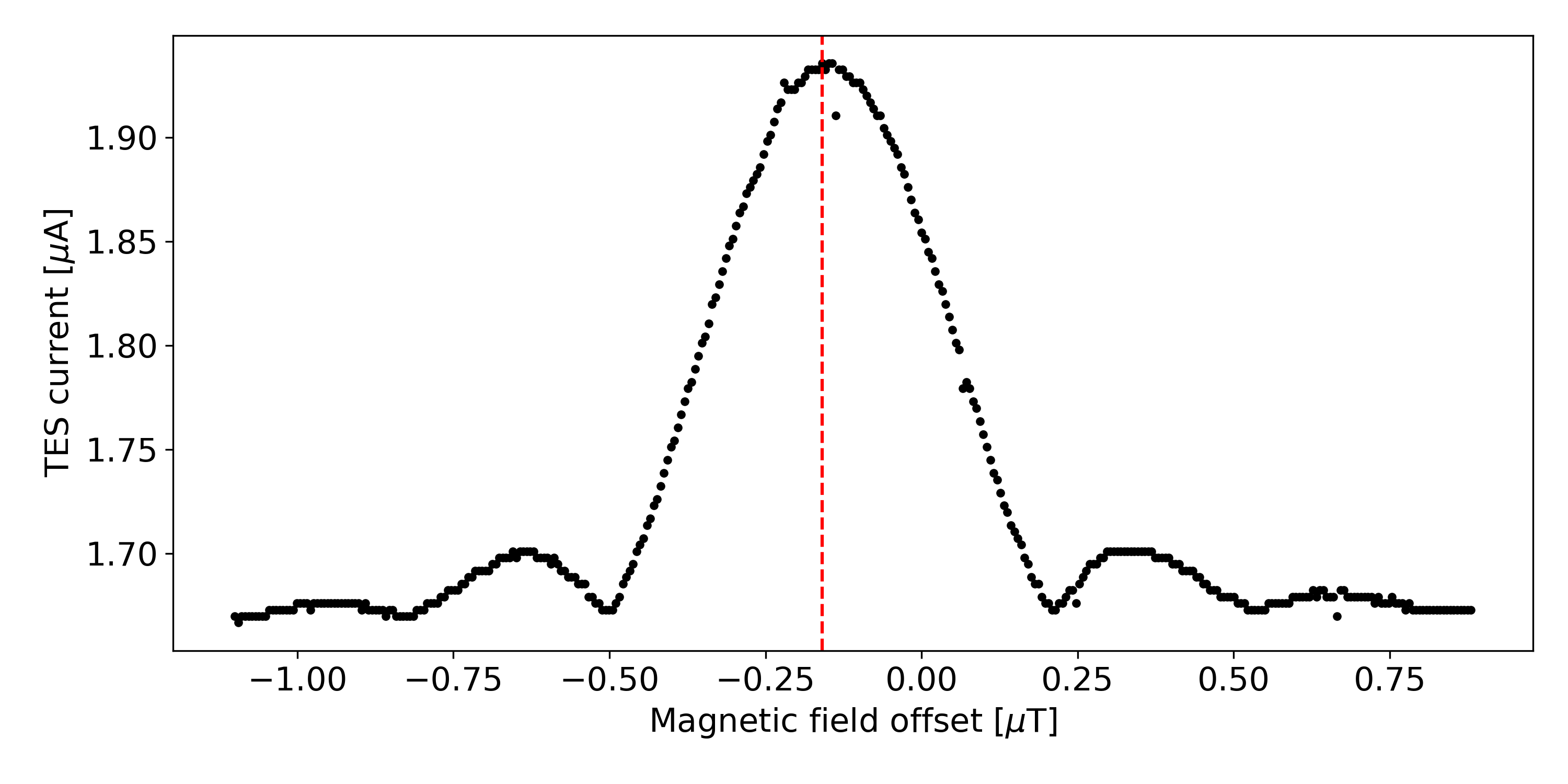}
\caption{Example of an $I\urss{TES}$~vs.~$B$ curve measured at a fixed temperature, close to $T\urss{C}$. The optimal magnetic field offset (red dashed line) is estimated from the peak of the Fraunhofer pattern, where $I\urss{TES}$ is maximum.}\label{fraunhofer}
\end{figure}

The operating assumption is that the best spectral performance is achieved when an external magnetic field equal to the residual field without self-field is applied. To estimate this value,  curves of $I\urss{TES}$ vs.  $V\urss{TES}$ at different magnetic fields $B\urss{ext}$ are measured to extract how the calibrated TES current $I\urss{TES}$ changes as a function of applied field for a given bias point. We then obtain $I\urss{TES}$ vs.  $B\urss{ext}$ curves, from which the well-known Fraunhofer-like trend can be extracted. The peak of such curve for different bias points can be used to extrapolate the cancelling field corresponding to when no current is flowing through the TES. A more detailed description of this method can be found in our previous report\cite{srontdm}.

As a sanity check, we also estimated the optimal cancelling field with an alternative method, raising the bath temperature just below $T\urss{C}$ and measuring the $I(B)$ curve with the minimum bias voltage to bring the TES in transition. In this configuration,  $I\urss{TES}$ is low and the self-field can be considered negligible, so that the peak of the measured Fraunhofer pattern can be assumed to be purely due to the residual magnetic field. An example of this measurement is reported in Figure~\ref{fraunhofer}.

\begin{figure}[h]%
\centering
\includegraphics[width=0.99\textwidth]{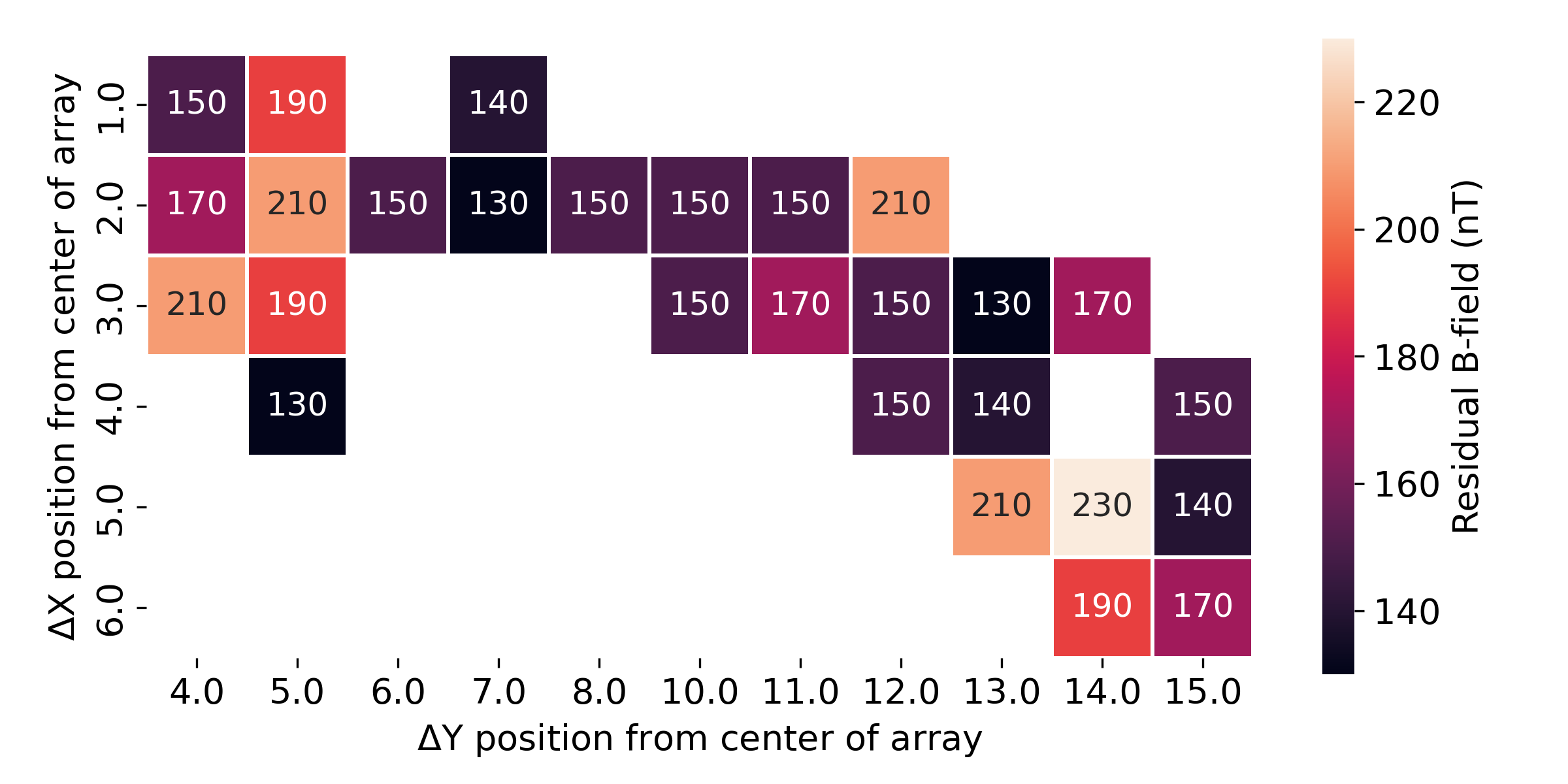}
\caption{Map of the canceling magnetic field measured in the TES array for one TDM column. \textcolor{black}{The axis labels represent the distance on the TES array plane in pixel number units.}}\label{bfieldmap}
\end{figure}

We found the average of the cancelling fields measured with both methods to be consistent.  We measured an average cancelling magnetic field at a level of 150~nT with no evident gradient across the array, with a few TESs showing a variation in residual field with respect to neighbouring pixels (see Figure~\ref{bfieldmap}).  The reason for this variation is not identified, but it could be due to either uncertainty in the measurement or a real effect in the TES array due to $e.g.$ non-perfect uniformity.  For reference, this value is nevertheless within the $\Delta B < 1 \upmu$T requirement for X-IFU for magnetic-field gradient across the TES array. Furthermore, this value of average residual field is comparable with that typically measured for other TES setups read out under FDM hosted in the same cryostat \cite{fdm}. The minimization of magnetic field gradients is important for flight operation,  because TES pixels within a TDM readout column will share their dc-voltage bias and thus cannot be tuned individually to compensate for eventual different magnetic-field offsets.

\subsection{Bandwidth optimization}

As mentioned in Section~\ref{secflex}, the old flex design resulted in a large parasitic inductance which caused underdamped oscillations (``ringing") in the electrical signals of the systems. Such ringing is expected to be initiated by two mechanisms: the switching transients sequentially activating the TDM rows and large changes in the feedback signal, such as when the readout switches to/from a row reading out a TES hit by an X-ray. 

\textcolor{black}{This mechanism can be characterized by injecting to the feedback coil of one SQ1 (first-stage SQUID) a signal of amplitude comparable to the one caused by an X-ray releasing energy in the TES. For this system, the pulse height for a 6~keV photon roughly corresponds to a signal of 2$\Phi_0$, with $\Phi_0 \approx 2.067\cdot10^{-15}$~Wb being the magnetic flux quantum and representing a single period in the $I\urss{SQ1}$~vs.~$\Phi\urss{SQ1}$ characteristic curve. }

\begin{figure}[!h]%
\centering
\includegraphics[width=0.99\textwidth]{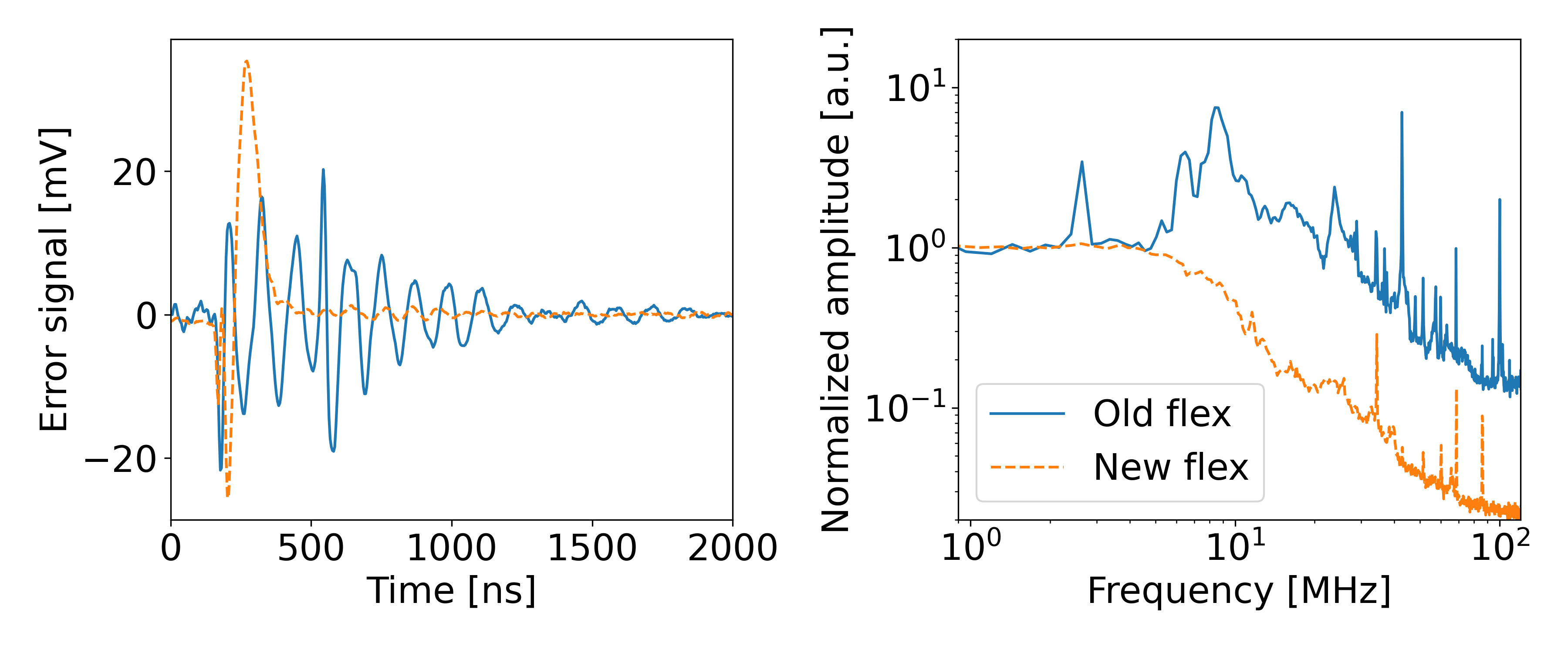}
\caption{Signal measured at the ADC input after injecting a 2$\Phi_0$ step function at one SQ1 feedback coil.  For reference, the full modulation depth of a combined SQ1/SSA SQUID curve is typically at a level of 200 mVpp. The old flex design (blue) resulted in a ``ringing" with an extended settling time, while the new flex design (dashed orange) in a transient with large amplitude and short decay time.}\label{ringing}
\end{figure}

This experiment was performed first for the old flex and the observed response is plotted in blue in Figure~\ref{ringing}. We measured a peak amplitude is such oscillation at a frequency of $\approx 8$~MHz, well within the system bandwidth. In TDM systems with a regular short harness, such ringing is present but at a higher frequency that is outside the typical 12-15~MHz readout bandwidth, allowing for 160~ns row time readout.

With the old flex harness, the system performance could be retained by operating with an increased row time of 360~ns, corresponding to a line rate of $\approx$~2.7 MHz and sampling for 64~ns instead of 32~ns.  As a consequence, to maintain roughly the same frame time, the multiplexing factor was reduced from 32 to 16. With this configuration, both at NIST and at SRON, similar systems were able to achieve an average energy resolution of 2.8 eV FWHM \cite{srontdm}.

Subsequently, the new flex design was implemented in the system. A ringing characterization measurement was repeated, by injecting a 2$\Phi_0$ signal at the SQ1 feedback coil.  In this case (Figure~\ref{ringing}, dashed orange curve), no underdamped oscillations were observed. However, most likely due to the larger capacitance to ground in the new design, a significantly reduced bandwidth was observed, at the level of $\approx$~5~MHz.

In consequence of this,  the decay time of the transient was at a level of 200~ns, longer than the 160~ns row time, which made the use of the nominal timing settings infeasible. This effect on the switching transient due to finite analog bandwidth is known and consistent with what is reported in Durkin \emph{et al.} (2021) \cite{durkinbw},  and it is equivalent to applying an aliasing filter to the SQ1 feedback signal path.

\begin{figure}[!h]%
\centering
\includegraphics[width=0.9\textwidth]{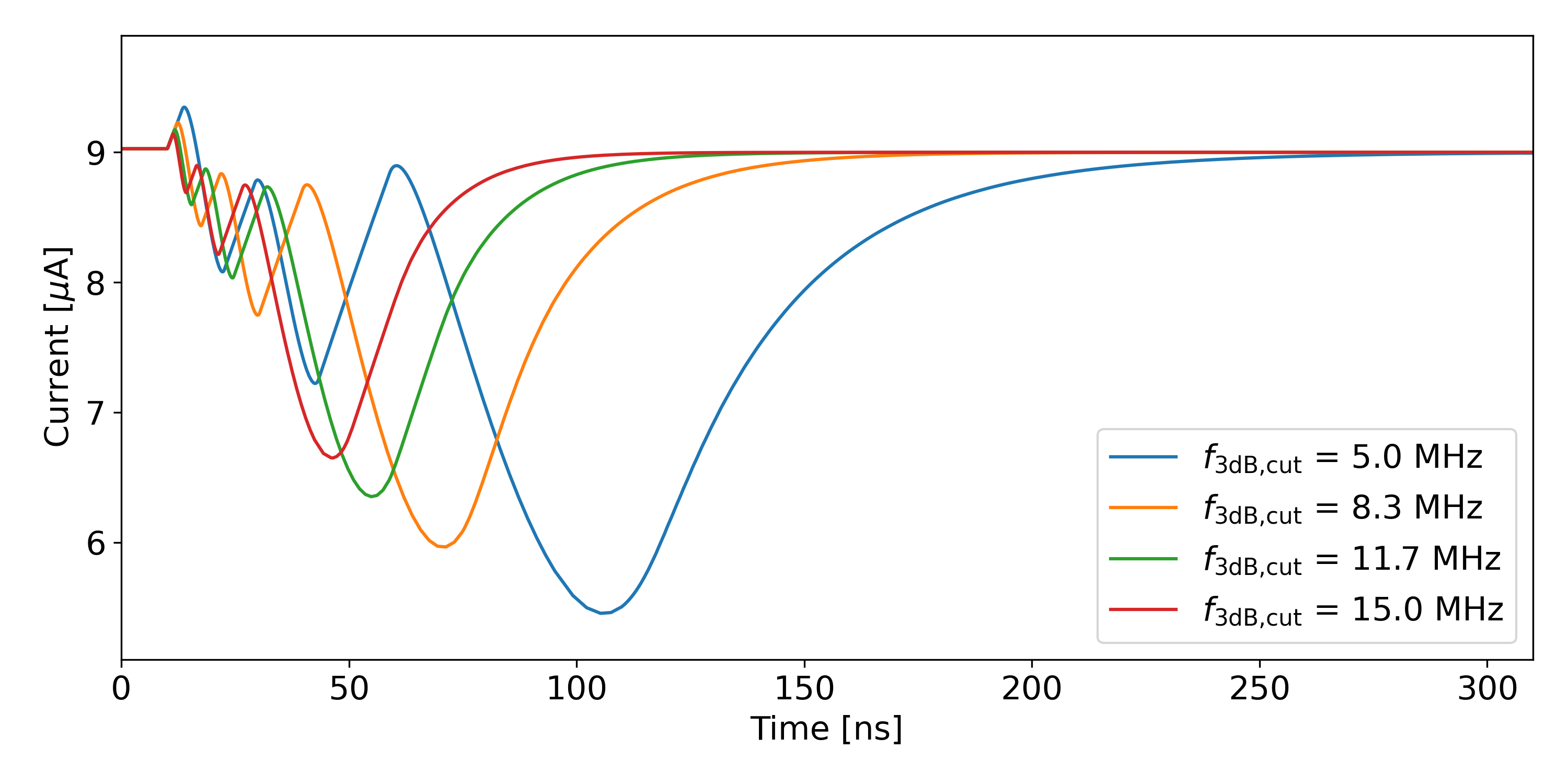}
\caption{Simulated signal at the input of the ADC as a function of a cut-off frequency in the SQ1 feedback path. The simulation is performed using a SQUID look-up table model in the Dymola, with a 3$\Phi_0$ step function at the feedback of a single SQ1 in FLL and no input signal at the TES. \textcolor{black}{Assessment method reprised from Durkin \emph{et al.} (2021)\cite{durkinbw}.}}\label{bwsim}
\end{figure}

We further verified this by simulating the system behavior in a Dymola\cite{dymola} model. A low pass filter was inserted in the feedback path to simulate the limited bandwidth of the flexes and the $f\urss{3dB}$ cut-off was varied to several values. The impact on the transients is reported Figure~\ref{bwsim}, from which a cut-off at a level of 5~MHz is qualitatively consistent with the experimental observations.

\begin{figure}[!h]%
\centering
\includegraphics[width=0.9\textwidth]{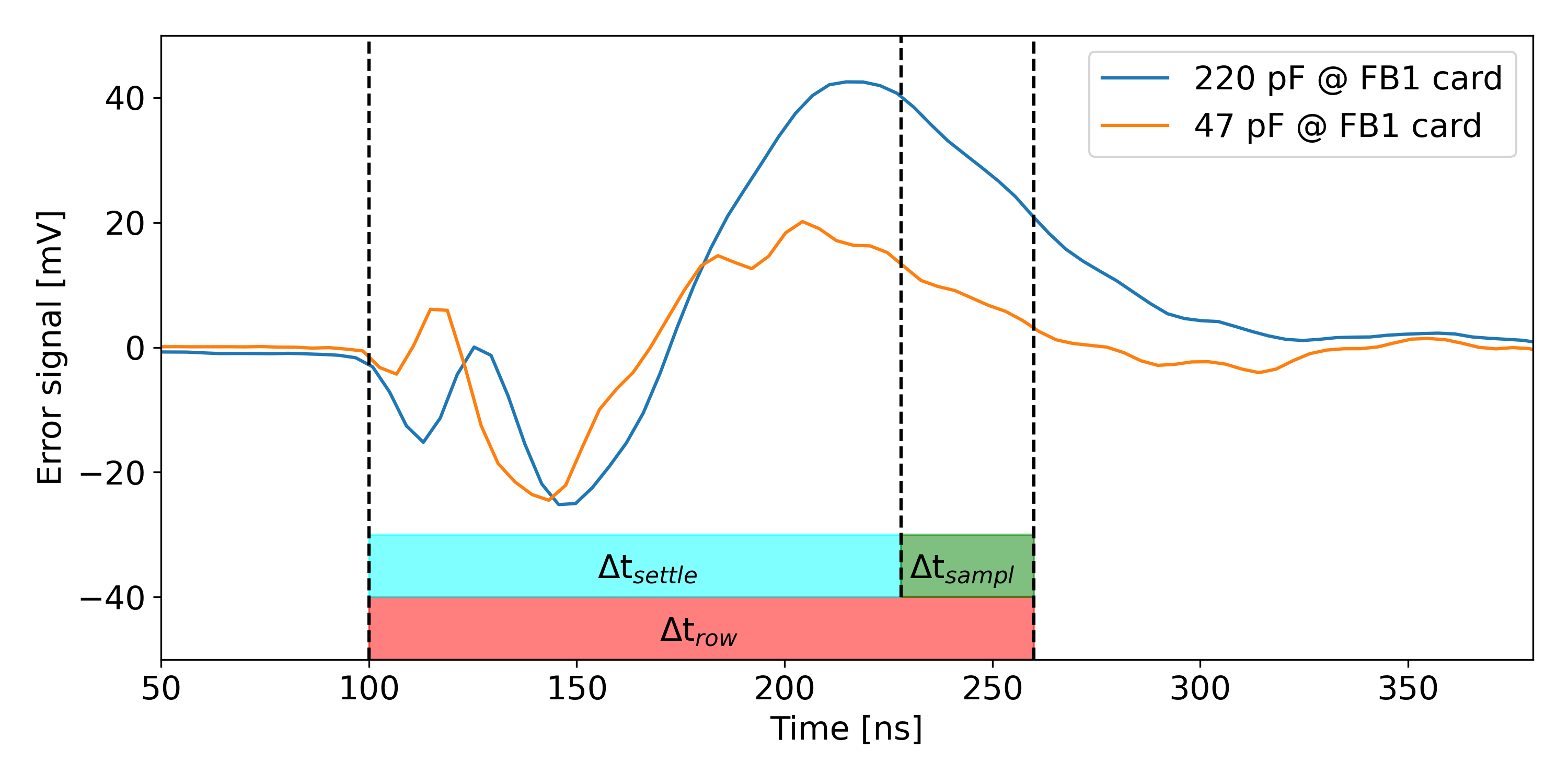}
\caption{Comparison between the switching transient obtained with a 2$\Phi_0$ signal injected \textcolor{black}{at the SQ1 feedback coil (FB1)}, before (220 pF) and after (47 pF) the modification to the Tower card. \textcolor{black}{For 160 ns row time, the 32~ns sampling window falls within the tail of the transient in both cases, however with the 47~pF capacitor it is sufficiently stable and close to the baseline value to allow for 32-row operation.  Assessment method reprised from Durkin \emph{et al.} (2021)\cite{durkinbw}.}}\label{time_transient}
\end{figure}

To mitigate this effect, we modified the analog feedback circuitry in the Tower, increasing the cut-off of the RC bandpass filter by modifying the value of the capacitor to ground from 220~pF to 47~pF. The impact on the system is reported in Figure~\ref{time_transient}. In this case, we did observe an increase in bandwidth,  at a level of 7 MHz. The amplitude of the transient decreased by a factor 2 and the decay time reduced to less than the 160~ns of the required row time.  We considered this feasible for practical operation, re-tuning the configuration for settling time $\Delta t\urss{settle}$ and sampling time $\Delta t\urss{sample}$ as described in the caption of the figure. This configuration was then used to carry out the X-ray performance measurements described in the next Section.

\section{System performance demonstration}\label{demo}

To measure the X-ray performance of the system, about 10,000 X-ray events per pixel are collected from the Fe-55 source installed on the snout's chimney, with a count rate of approximately 1.2 counts per second per pixel. The energy of the photons is estimated using the optimal filtering technique\cite{optfit}. To assess the spectral performance, the MnK$\upalpha$ model \cite{holzer} is fitted to the collected, calibrated data.  The model is convolved with a Gaussian broadening to parametrize the FWHM energy resolution of the detectors. The goodness of the fit is estimated with the maximum likelihood method by means of a $\chi^2$ statistics modified to mitigate biases due to bin size or number of events\cite{fowler}. 

Each channel is activated by biasing the SSA, SQ1 and TES detectors on their optimal operating point. For the SSA, this is done by measuring its V-$\Phi$ curves for different bias currents and selecting the one with the largest modulation depth.  The flux offset is chosen to operate the SSA on the steeper slope of its $V\urss{out}$~vs.~$\Phi$ curve, both to minimize the system noise and to maximize the readout bandwidth. The same is repeated for the bias current of the SQ1s, selecting the one producing the maximum modulation depth. The flux offset of the SSA is then varied iteratively per each row to find the value maximising each SQ1's modulation depth. A voltage offset is then applied to center the FLL signal around zero, to make optimal use of the feedback DAC dynamic range.

\begin{figure}[h]%
\centering
\includegraphics[width=0.95\textwidth]{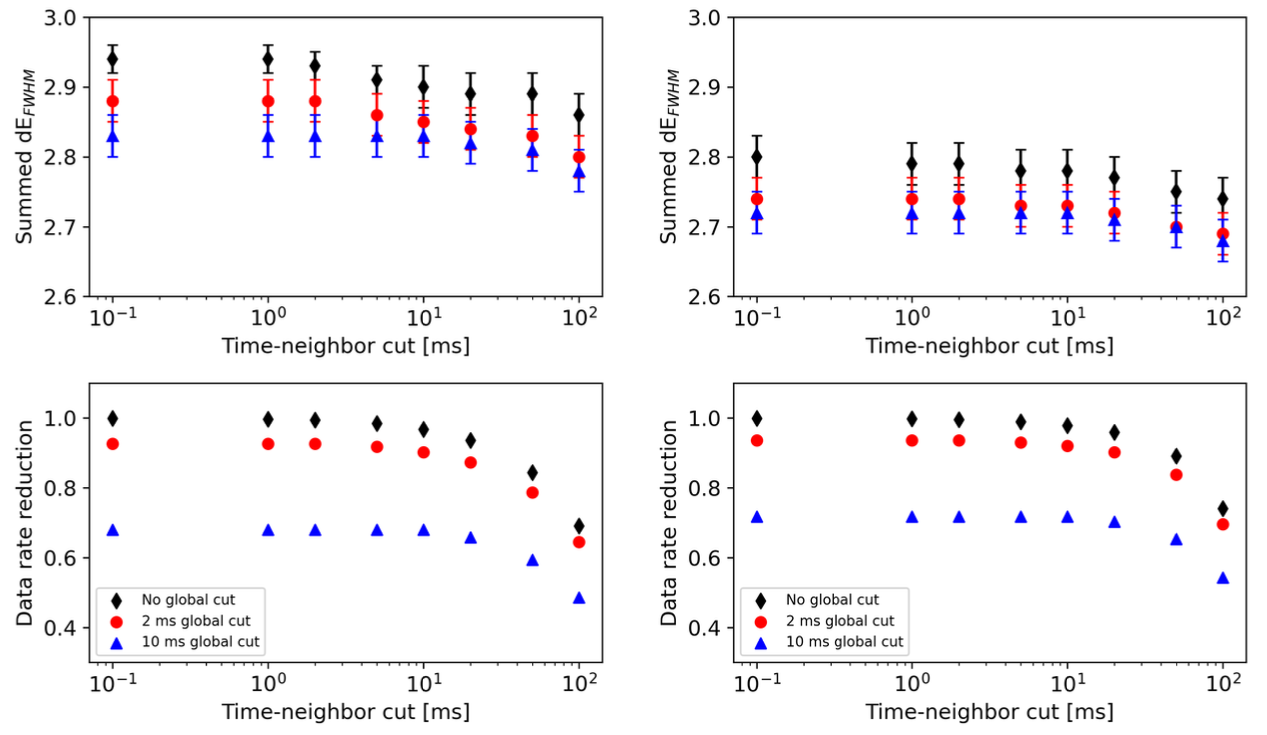}
\caption{Impact on spectral performance of crosstalk rejection in the data analysis pipeline for the two count rates used. }\label{xtalk}
\end{figure}

The TES bias is set to a value corresponding to roughly 17\% of its normal resistance value,  which during early measurements had shown to yield on average the best spectral performance, at the 150~nT magnetic field offset. 

Measurements performed in this configuration showed a co-added energy resolution of 2.95~$\pm$~0.03~eV.  Implementing electrical crosstalk rejection in the analysis pipeline,  based on known and understood crosstalk mechanisms in TDM systems\cite{xtalkdurkin,durkin2019}, showed an improvement on the co-added performance.  In particular, we applied time-neighbour and global cuts, as summarised in Figure~\ref{xtalk}. 

Time-neighbor cuts tag events which happen within a certain time window from an event happening in a subsequent or preceding row, while global cuts tag events happening within a time window regardless of their row number. In both cases, the cuts are limited to within the same Column. To simplify the data analysis pipeline, we have applied here a crosstalk-rejection strategy that cuts all perpetrator-victim pairs within a readout Column, with a time window of 2~ms.  This strategy is suboptimal with respect to what is being envisaged for X-IFU, so that the throughput of surviving counts in this data set should not be understood as an analysis of what is possible for the flight pipeline of X-IFU. Furthermore, the impact of crosstalk on in-flight performance is expected to be lower than what is measured in the laboratory with the approximately monochromatic Fe-55 source,  since the typical models assume a Crab-like energy spectrum,  with an average energy lower than 6~keV \cite{xtalkdurkin, smith2020}.

Experiments with a 1.2 cps count rate showed a larger impact on the co-added system performance. However, typical TDM laboratory demonstrations use a count rate at a level of 0.3 cps per pixel, similar to the expected average count rate for X-IFU when not observing high intensity sources such as supernova remnants, gamma ray burst afterglows or XRBs (X-ray binaries) in our galaxy. We therefore repeated the experiment by reducing the count rate to about 0.7~cps per pixel, by increasing the thickness of the Al foil in front of the cold $^{55}$Fe source.  

\begin{figure}[h]%
\centering
\includegraphics[width=0.99\textwidth]{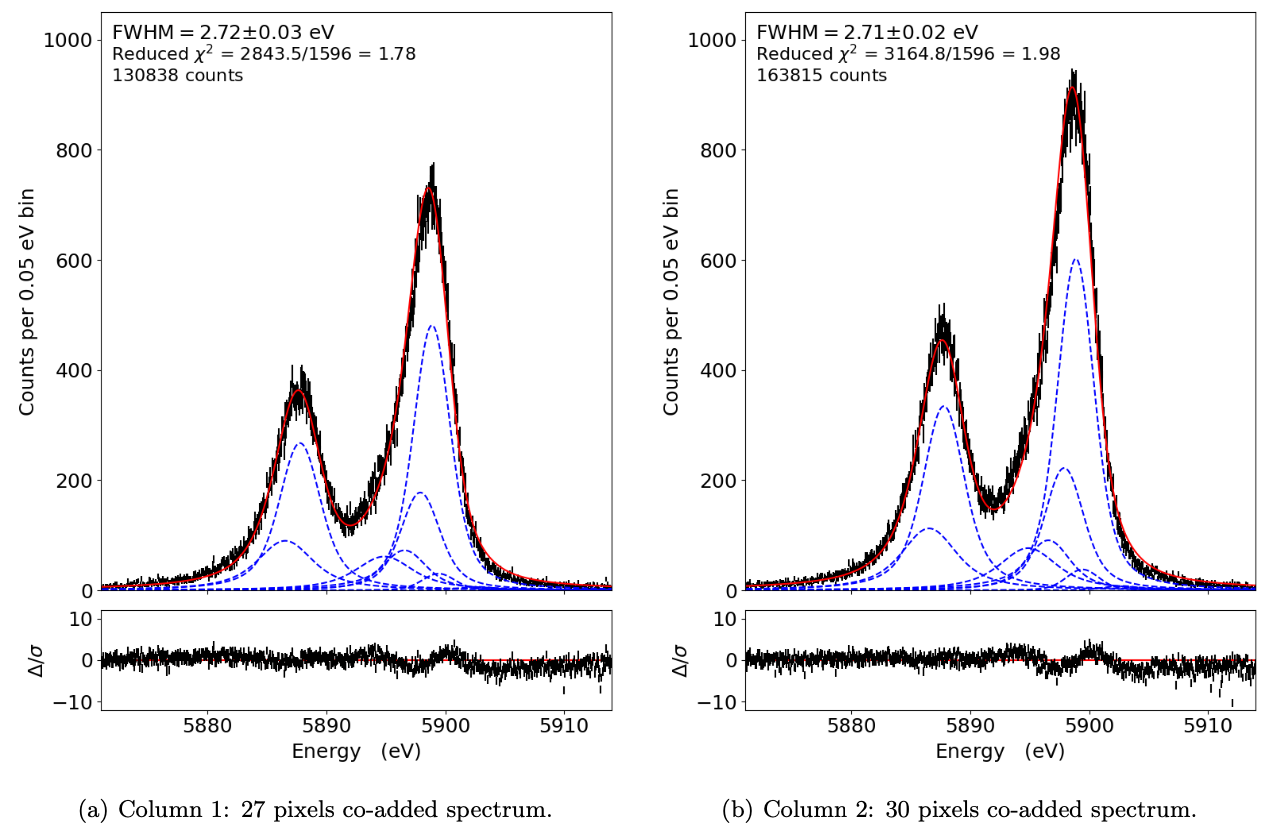}
\caption{Co-added spectra measured for the two Columns of the system in 32-row readout: the histogram with the collected data in black, the best fit of MnK$\upalpha$ model convolved with a Gaussian broadening in red,  the individual spectral components of the model in dashed blue.  The bottom panels show the residuals of the fit. In Column 1 (a) 5 pixels were excluded from the analysis because of either malfunctioning components in the MUX or TES chip, or TESs receiving significantly fewer counts because of aperture mask mis-alignment. In Column 2 (b) 2 pixels were excluded from the analysis, for the same reasons. The excess residual on the high-energy tail of the MnK$\upalpha_1$ line is interpreted as due to the addition of spectra from pixels with different energy resolution.}\label{results}
\end{figure}

With the typical crosstalk cuts for laboratory TDM systems (20 ms window for time-neighbour and 2 ms window for global) we obtain a co-added performance of $2.72\pm0.03$~eV, with a data rate reduction of $\approx$~8\% which already well fits the throughput requirements for X-IFU \cite{xifu}.  From this analysis 5 of the 32 rows are excluded, due to either faulty cold components (TES or SQ1) or TES receiving significantly fewer counts because of aperture mask misalignment.

With the same configuration, we repeated the experiment for the second Column of the setup. We measured consistent performance, with a co-added energy resolution of $2.71\pm0.02$~eV. For this Column, 2 pixels were excluded from the analysis for the same reasons listed above. 

The results for both Columns are summarised in Figure~\ref{results}.

\section{Conclusions and future outlook}\label{concl}

We reported on the characterization of a cryogenic test-bed to verify the performance of a TDM system with long harness, towards the integration of the new Focal Plane Assembly Development Model for X-IFU.

The re-designed flex harness allowed to mitigate the ringing observed in the previous version, at the cost of a reduced analog bandwidth.  A bandwidth of $\approx$~7~MHz was obtained by modifying the FB1 board in the analog electronics and choosing different settings for the ADC low pass filter in the DRE.  In this way, it was possible to operate the system with a performance at the nominal level, with an energy resolution measured at a level of 2.7~eV at 5.9~keV energy.

This demonstration represents a fundamental progress for the FPA-DM1.1 of X-IFU as it is the first time that a TDM system is operated with a 1.5~m long harness, closer to the X-IFU's 2.3~m requirements. 

NewAthena's instrument will employ twisted-pair looms in balanced differential operation, and so will the FPA-DM1.1.  \textcolor{black}{Having reached the expected level of performance for the system, in the near future we will replace the current flex harness with twisted-pair looms, to assess their impact not only on performance but also on crosstalk and analog bandwidth. This will be an important intermediate step before the integration and test campaign of the X-IFU FPA-DM1.1,  foreseen for Q2 2025, which will allow to assess the TDM readout operation in a configuration more representative of X-IFU. }The comparison of the system behavior with a flex harness versus twisted-pair looms will be reported on a separate paper.

\section*{Acknowledgements}

SRON is financially supported by the Nederlandse Organisatie voor Wetenschappelijk Onderzoek.
This work is part of the research programme Athena with project number 184.034.002, which is (partially) financed by the Dutch Research Council (NWO).

The material is based upon work supported by NASA under award number 80GSFC21M0002.

\section*{Data availability}

The corresponding author makes available the data presented in this paper upon reasonable request.

\section*{Conflict of interest}

The authors declare there are no financial interests, commercial affiliations, or other potential conflicts of interest that have influenced the objectivity of this research or the writing of this paper.

\newpage

\section*{List of Figure captions}

\begin{itemize}

\item Figure 1: Cryogenic components of the setup installed in the dilution refrigerator: 50 mK snout in Al superconducting shield (A),  Fe-55 X-ray source mounted on a Al chimney in front of the TES array (B), Cu aperture mask placed in between the TES array and the X-ray source (C), two-stage Kevlar suspension system (D), 4 K board (E), SSA SQUIDs (F), inter-thermal stage boards (G) for the NbTi braided looms connecting to the 4 K to the 50~mK snout, flex harness (H), Tower (I) during system assembly,  DRE (J) mounted in an electronics rack.

\item Figure 2: Cross-section of the flex harness, comparing between the old (a) and current (b) flex design.  The direction of current flow along the conductors is into and out of the plane of the paper.  The pair pitch between adjacent wire pairs (not shown here) is 2.54~mm. The differences stand in the conductor material, symmetry between supply and return line and their individual width.

\item Figure 3: Example of an $I\urss{TES}$~vs.~$B$ curve measured at a fixed temperature, close to $T\urss{C}$. The optimal magnetic field offset (red dashed line) is estimated from the peak of the Fraunhofer pattern, where $I\urss{TES}$ is maximum.

\item Figure 4: Map of the canceling magnetic field measured in the TES array for one TDM column.  The axis labels represent the distance on the TES array plane in pixel number units.

\item Figure 5: Signal measured at the ADC input after injecting a 2$\Phi_0$ step function at one SQ1 feedback coil.  For reference, the full modulation depth of a combined SQ1/SSA SQUID curve is typically at a level of 200 mVpp. The old flex design (blue) resulted in a ``ringing" with an extended settling time, while the new flex design (dashed orange) in a transient with large amplitude and short decay time.

\item Figure 6: Simulated signal at the input of the ADC as a function of a cut-off frequency in the SQ1 feedback path. The simulation is performed using a SQUID look-up table model in the Dymola, with a 3$\Phi_0$ step function at the feedback of a single SQ1 in FLL and no input signal at the TES.  Assessment method reprised from Durkin \emph{et al.} (2021)\cite{durkinbw}.

\item Figure 7: Comparison between the switching transient obtained with a 2$\Phi_0$ signal injected at the SQ1 feedback coil (FB1), before (220 pF) and after (47 pF) the modification to the Tower card.  For 160 ns row time, the 32~ns sampling window falls within the tail of the transient in both cases, however with the 47~pF capacitor it is sufficiently stable and close to the baseline value to allow for 32-row operation.  Assessment method reprised from Durkin \emph{et al.} (2021)\cite{durkinbw}.

\item Figure 8: Impact on spectral performance of crosstalk rejection in the data analysis pipeline for the two count rates used.

\item Figure 9: Co-added spectra measured for the two Columns of the system in 32-row readout: the histogram with the collected data in black, the best fit of MnK$\upalpha$ model convolved with a Gaussian broadening in red,  the individual spectral components of the model in dashed blue.  The bottom panels show the residuals of the fit. In Column 1 (a) 5 pixels were excluded from the analysis because of either malfunctioning components in the MUX or TES chip, or TESs receiving significantly fewer counts because of aperture mask mis-alignment. In Column 2 (b) 2 pixels were excluded from the analysis, for the same reasons. The excess residual on the high-energy tail of the MnK$\upalpha_1$ line is interpreted as due to the addition of spectra from pixels with different energy resolution.

\end{itemize}

\end{document}